\documentclass[final,12pt,3p]{elsarticle} 
\usepackage{amssymb,amsthm,amsmath,amsfonts,graphicx,xcolor}
\usepackage[colorlinks=true]{hyperref}
\usepackage{caption}
\usepackage{tabularx}
\usepackage{array}
\usepackage{subfigure}
\usepackage{float}
\usepackage[noend]{algpseudocode}
\usepackage{algorithmicx,algorithm}
\usepackage{indentfirst} 

\graphicspath{{Myfigfiles/}} 
\biboptions{numbers,sort&compress}
\journal{Chaos}

\begin{document}
\begin{frontmatter}

\title{Solving Fokker-Planck equation using deep learning}

\author[firstaddress,secondaddress]{Yong Xu\corref{mycorrespondingauthor}}
\ead{hsux3@nwpu.edu.cn}
\cortext[mycorrespondingauthor]{Corresponding author at: Department of Applied Mathematics, Northwestern Polytechnical University, Xi'an 710072, China.}
\author[thirdaddress,firstaddress]{Hao Zhang}
\author[forthaddress]{Yongge Li}
\author[firstaddress]{Kuang Zhou}
\author[firstaddress]{Qi Liu}
\author[fivethaddress,sixthaddress]{J{\"u}rgen Kurths}

\address[firstaddress]{Department of Applied Mathematics, Northwestern Polytechnical University, Xi'an 710072, China}
\address[secondaddress]{MIIT Key Laboratory of Dynamics and Control of Complex Systems, Northwestern Polytechnical University, Xi'an 710072, China}
\address[thirdaddress]{Department of Engineering Mechanics, Northwestern Polytechnical University, Xi'an 710072, China}
\address[forthaddress]{Center for Mathematical Sciences and School of Mathematics and Statistics, HuazhongUniversity of Science and Technology, Wuhan 430074, China}
\address[fivethaddress]{Potsdam Institute for Climate Impact Research, Potsdam 14412, Germany}
\address[sixthaddress]{Department of Physics, Humboldt University Berlin, Berlin 12489, Germany}

\begin{abstract}
The probability density function of stochastic differential equations is governed by the Fokker-Planck (FP) equation. A novel machine learning method is developed to solve the general FP equations based on deep neural networks. The proposed algorithm does not require any interpolation and coordinate transformation, which is different from the traditional numercial methods. The main novelty of this paper is that penalty factors are introduced to overcome the local optimization for the deep learning approach, and the corresponding setting rules are given. Meanwhile, we consider a normalization condition as a supervision condition to effectively avoid that the trial solution is zero. Several numerical examples are presented to illustrate performances of the proposed algorithm, including one- and two-dimensional systems. All the results suggest that the deep learning is quite  feasible and effective to calculate the FP equation. Further, influences of the number of hidden layers, the penalty factors, and the optimization algorithm are discussed in detail. These results indicate that the performances of the machine learning technique can be improved through constructing the neural networks appropriately.
\end{abstract}


\end{frontmatter}


The Fokker-Planck (FP) equations, as an indispensable part of stochastic dynamics, have been widely used in many different fields, such as physics, chemistry, biology, etc. The probability density functions can be obtained from the associated FP equations, which play an extremely important role in stochastic resonance, mean first passage time, engineering reliability, etc. Generally speaking, however, the exact solutions of the FP equations can only be obtained under some strict conditions. Thus, a number of numerical methods like finite element method, path integral technique, etc., has been proposed to solve the FP equations. These methods inevitably require mesh or associated transformations, which increases the amount of computation and operability. The problem becomes worse when the system dimension increases. Only recalculation or interpolation can be used when it is necessary to solve the value of points not on the grid. In this paper, we develop a novel deep learning method which well fits the nonlinear function to solve the FP equation. We add the normalization condition to avoid the trial solution to be zero. Meanwhile, we design a penalty factor for the loss functions to avert the appearance of local optimum. Furthermore, a detailed algorithm procedure is described and employed to four illustrative examples including one- and two-dimensional systems. The results illustrate the validity and maneuverability of the developed machine leaning to solve the FP equation compared with the exact or numercial results. 

\section{Introduction} 
The FP equation describes the time evolution of the probability density functions (PDFs) of stochastic complex systems which is a quantitative equation changing a stochastic problem to a deterministic one. It is a central topic in stochastic issues due to it's important relation to PDFs, stochastic resonance \cite{wang2016levy}, mean first pass time \cite{li2016levy} and diffusion, et. al. Besides, it has gained broad applications in statistical physics \cite{Escobedo1999Radiation}, chemistry \cite{Ceccato2018Remarks,Grafov2015Fokker}, biology \cite{Karcher2006A,xu2013levy}, engineering \cite{Fokou2016Probabilistic}, economy and finance \cite{Furioli2016Fokker}. Its exact stationary solution can only be obtained for few conceptual models under very strict conditions, and it is difficult to find its exact stationary solution in general \cite{J1996The}. Therefore, several numerical techniques have been developed to achieve solving of the FP equations \cite{N2014Finite,Gal2007Stochastic,Jiang2015A,Sepehrian2015Numerical,Drozdov1998Accurate,Chandrika2009Path,Zan2019Path,Biazar2010Variational,Torvattanabun2011Numerical,Sun2017A,Zheng2010A}.

In general, the PDFs of stochastic dynamics can be obtained from two main categories: One is directly to solve the FP equation through several techniques like finite element method \cite{N2014Finite,Gal2007Stochastic}, finite difference method \cite{Jiang2015A,Sepehrian2015Numerical}, path integral method \cite{Drozdov1998Accurate,Chandrika2009Path,Zan2019Path}, variation method \cite{Biazar2010Variational,Torvattanabun2011Numerical} and Galerkin method \cite{Sun2017A,Zheng2010A}. These approachs focus on discretizing the computational domain into a set of grid points and solving approximate solutions on the grid, but the solution is only calculated at the grid point, and the evaluation of any other point requires an interpolation or other reconstruction techniques. For one-dimensional (1D) problems, these methods are very effective, but in two-dimensional (2D) and higher dimensional systems, they consume much computing resources and even have troubles to overcome the curse of dimensionality. In addition, a sparsity of the grid seriously affects the accuracy of calculation, while a dense grid will inevitably lead to a sharp increase in the amount of calculation. The other category is solving the stochastic differential equations and then statistics its transition probability density, the typical one is the so-called  Monte Carlo method \cite{Norio2013Thermal,Hirvijoki2015Monte}. Although this approach does not require interpolation, it needs a large number of sample paths and the accuracy of its solution is related to the amount of generated data. For a 2D system only, the computational cost of solving the joint probability density function (JPDF) may be too large to be obtained. Therefore, it is still a problem that needs to be solved to find an alternative technique without interpolation calculation and with rather low computational complexity. The use of artificial neural networks (ANNs) provides a promising way to solve this problem. 

Accually, ANNs have attained a considerable attention as robust and effective tools for the function approximation \cite{Barrow2018The,Schirmer2017Statistical}. In 1989, Cybenko \cite{Cybenko1989Approximation} theoretically proved that an ANN using sigmoid as the activation function could fit any continuous function over a compact subset of $n$-dimensional real space. A large number of scholars has focused on using ANNs to solve ordinary differential equations and parital differential equations due to the strong ability of ANNs to fit continuous functions \cite{Lagaris1998Artificial,Leephakpreeda2002Novel,van1995Neural,Beidokhti2009Solving,Malek2006Numerical,Berg2017A,RaissiPhysics,Raissi2017Physics1,Raissi2017Physics2,Han2017Solving,Sirignano2017DGM}. Here a supervised learning model was set to discretize the computation interval and boundary interval as the input of the network, and took the form of differential equations and boundary conditions as the supervised conditions of the network training, so as to make the trial solution approach the true solution through a minimization criterion. From the perspective of the network construction, it can be divided into two forms: a single hidden layer neural network and a multiple hidden layer neural network. To estimate the partial derivatives in  differential equations, the classification can be divided into three categories. A single hidden layer neural network is used to directly apply the derivative formula \cite{Lagaris1998Artificial,Leephakpreeda2002Novel,van1995Neural,Beidokhti2009Solving,Malek2006Numerical}, while others \cite{Berg2017A,RaissiPhysics,Raissi2017Physics1,Raissi2017Physics2} constructed the network using a multi-layer neural network with an automatic differential technique to estimate the derivative, and in \cite{Sirignano2017DGM} the Monte Carlo method is taken to estimate the derivatives under the guidance of Galerkin's idea. To summarize, solving the differential equations with ANNs technique offers the following preferences in comparison with classical numerical schemes: (1) ANNs learn analytic solution. (2) Solution search does not require any coordinate transformation. (3) As the number of sampling points increases, the computational complexity does not increase rapidly. (4) Calculate the value of the solution quickly. 
 
The FP equation is a kind of second-order parabolic parital differential equations, but the above methods cannot be applied to solving the FP equation. There are two main reasons: first there is no constant term in the format of FP equation and the boundary condition is zero, which makes zero become the trial solution during the training which is not the exact solution. So it is necessary to increase the supervision condition to solve this problem. The second is that the traditional mean square error loss function can not be used to train the network effectively due to the increased monitoring conditions, so it is necessary to change the format of the loss function. 

In this paper, we propose a method applying deep learning to solve the FP equation to overcome the mentioned points. In addition to take the form of the equation and the boundary condition as the supervision condition, we will also take the normalization condition as the supervision condition. Hence the solution of the FP equation is a PDF or JPDF, which can avoid the situation that the trial solution is zero. In order to train the three components of the loss function well, a penalty factor is introduced to solve this problem.

The structure of this paper is as follows: Section 2 gives a detailed description of the method to solve a FP equation using deep learning which is called the DL-FP method, and the proposed algorithm scheme is presented. In Section 3, four examples consisting of 1D and 2D systems are analyzed by means of the DL-FP method, and the results are discussed. In Section 4, we will discuss three concerns in the algorithm processdure: 1. The effect of how to set the number of hidden layer on the calculated result. 2. The importance of the penalty factor in the algorithm. 3. How to select the optimization technique in the DL-FP algorithm. Finally conclusions are presented to close this paper.

\section{Methodology} 
\subsection{The FP equations}
Consider the following general $n$-dimensional SDE
\begin{equation}     
	\dot{X}=f(X)+\Lambda \Gamma, \label{eq:1}
\end{equation}
where $X=[x_1,x_2,...,x_k,...,x_n]^T$ is an $n$-dimensional variable $(n\geq 1)$, $\Lambda$ is a square matrix which is assumed here to be constant, $\Gamma$ is a vector of uncorrelated standard Gaussian white-noise processes, and $f(X)$ is a general $n$-dimensional vector function of $X$. For the eq.~\eqref{eq:1}, the solution process is still a Markov one and the joint PDF (JPDF) of the stationary response satisfies the reduced FP equation as follows
\begin{equation}     
	-\sum_{k=1}^{n} \frac{\partial}{\partial x_{k}}\left[f_{k}({X}) p( X)\right]+\sum_{j=1}^{n} \sum_{k=1}^{n} \frac{\partial^{2}}{\partial x_{j} \partial x_{k}}\left[\frac{{m}_{j k}(X)}{2 !} p({X})\right]=0, \label{eq:2}
\end{equation}
where $m_{i j}, i,j=1,2,...,n$ are the elements of the matrix $M$ given by $M=2 \pi \Lambda \Lambda^{\mathrm{T}}$. $p(X)$ is the transition PDF of Eq.~\eqref{eq:1}, which must satisfies the normalization condition and $p({X})$ tends to zero as $X$ tends to infinity.

In order to describe Eq.~\eqref{eq:2} expediently, we can rewrite it as
\begin{equation}     
	\mathcal{N}\left(p(X), p_{X}(X), p_{X X}(X)\right)=0, \label{eq:3}
\end{equation}
where $\mathcal{N}[\cdot]$ is the FP differential operator.

\subsection{The DL-FP algorithm}
In this section, we will describe the numerical scheme to solve the FP equation via a deep ANN and we name it as DL-PF algorithm. 

\begin{figure}[H]
	\centering
	\includegraphics[width=0.6\textwidth]{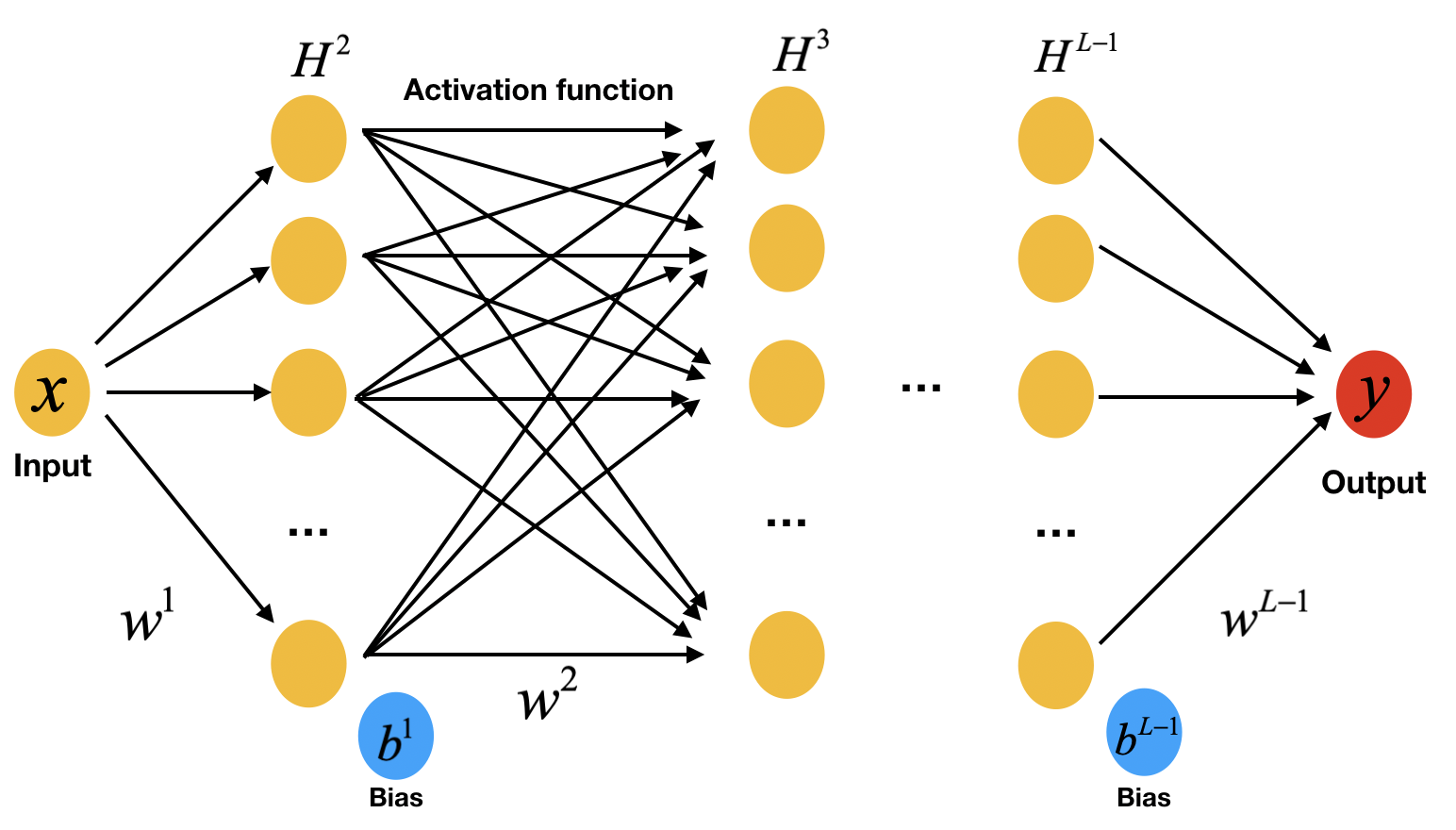}  
	\caption{The structure of a deep nerual network. \label{fig:1}}
\end{figure}

Fig.~\ref{fig:1} illustrates a typical structure of an ANN which is also called deep learning with $L$ layers, where $x$ is the input layer, and $y$ is the output one. $H^{l}=\left[h_{1}^{l}, h_{2}^{l}, h_{3}^{l}, \ldots, h_{n_{l}}^{l}\right](2 \leq l \leq L-1)$ is the output of the $l$-th hidden layer and $n_l$ is the number of nodes in each layer. $W=\left[w^{1}, w^{2}, \ldots, w^{l} \ldots, w^{L-1}\right]$ are the weights of network and $w^l$ represents the weight from the $(l-1)$-th layer to the $l$-th layer. The size of the matrix $w^{l}$ is $n_{l-1} \times n_{l}$. $Bias=\left[b^{1}, b^{2}, \dots, b^{l} \dots, b^{L-1}\right]$ is the bias of the layers and $b^l$ represents the bias of the $(l+1)$-th layer with the size $n_{l} \times 1$.
\par
Then we can describe a feed-forward process as
\begin{equation}     
	h_{i}^{2}=F\left(w_{i}^{1} {X}+b_{i}^{1}\right), \label{eq:4}
\end{equation}
\begin{equation}     
	h_{i}^{l}={F}\left(\sum_{j=1}^{{n_{l-1}}} w_{i j}^{l-1} h_{i}^{l-1}+b_{i}^{l-1}\right)(3 \leq l \leq L-1), \label{eq:5}
\end{equation}
\begin{equation}     
	y=\sum_{i=1}^{{n_{L-1}}} w_{i}^{L-1} h_{i}^{L-1}+b^{L-1}, \label{eq:6}
\end{equation}
where $F$ represents the activation function (like a sigmoid, tanh, ReLU, etc.), and the output can be described as $y=\mathcal{F}(x;\theta)$, where $\mathcal{F}$ is the composite function which is made of the activation function, and $\theta=[W,Bias]$ is denoted as the network parameters.

In general, the current methods solving the differential equations with ANN can be defined as follows: Assuming a differential equation with the initial condition or boundary condition to be
\begin{equation}     
\left\{\begin{array}{l}{\mathcal{D}(y({ X}))=0, { X} \in \Omega} \\ {\mathcal{B}(y({ X}))={ g}({X}), { X} \in \partial \Omega}\end{array}\right., \label{eq:7}
\end{equation}
where $\mathcal{D}[\cdot]$ is the differential operator which contains the derivatives with respect to $X$, and $\mathcal{B}[\cdot]$ is the boundary operator. The domain data $\mathcal{T}_{D}=\left\{X_{i}^{D} \in \Omega\right\}_{i=1}^{{ N_{D}}}$ and the boundary data $\mathcal{T}_{B}=\left\{X_{i}^{B} \in \partial \Omega\right\}_{i=1}^{{ N_{B}}}$ are discretized as input data of ANN, and the corresponding output is denoted as $\widehat{y}=\widehat{y}({ X} ; \theta)$, where $\theta$ are the parameters of the underlying ANN. The loss function is denoted as $\mathcal{L}=E_{1}+E_{2}$, and

\begin{equation}     
	E_{1}=\frac{1}{N_{D}} \sum_{i=1}^{{ N_{D}}}\left|\mathcal{D}\left(\widehat{y}\left({ X}_{i}^{D}; \theta\right)\right)\right|^{2}, 
\end{equation}
\begin{equation}     
	E_{2}=\frac{1}{N_{B}} \sum_{i=1}^{{ N_{B}}}\left|\mathcal{B}\left(\widehat{y}\left({ X}_{i}^{B}; \theta\right)\right)-{ g}\left({ X}_{i}^{B}\right)\right|^{2},
\end{equation}
in which $N_D$ and $N_B$ represent the number of data in the training set and boundary one, respectively. Then, an optimization method is employed to train the network. This kind of training belongs to the supervised learning. The loss parts $E_{1}$ and $E_{2}$ guarante that the training solutions satisfy the form of the given equation and the boundary conditions, respectively. 

However, this method is not feasible to deal with the general FP equation \eqref{eq:3}. The training results will be zero with only the form of the equation and the boundary conditions. Therefore, we add the normalization condition as the supervision condition to avoid this confusion. At the same time, due to the addition of a supervision item, the training results in high-dimensional case will be invalidated. To address this problem, penalty factors will be introduced. Next, the detailed procedure of the proposed DL-FP algorithm  will be described as follows.

{\romannumeral1}) The step one is to select the training set. Since the input of an ANN should be a finite set, the range of independent variables of the FP equation needs to be truncated. Considering the FP equation with equilibrium states, the probability is zero when the independent variable is far from these states. So we should select a sufficiently large interval. To be more accurate, we can use a Monte Carlo method to find the training interval.

After finding the training interval, we can get the training set  $\mathcal{T}_{D}=\left\{X_{i}^{D} \in \Omega\right\}_{i=1}^{{N_{D}}}$. It is a discretization on each dimension of independent variables, and the step length is {$\Delta t$}. In the 1D case, we assume the interval to be $[a, b]$, then $N_{D}=(b-a) / \Delta t$, and the number of data in the training set is $N_D$. In the 2D case, we assume the interval to be $[a, b] \times[a, b]$ and the number of data in the training set is $\left(N_{D}\right)^{2}$. The boundary set is $\mathcal{T}_{B}=\left\{X_{i}^{B} \in \partial \Omega\right\}_{i=1}^{{N_{B}}}$  which describes the boundary of the training set. In the 1D case, the number of data in the boundary set is $N_B=2$, and it is $N_{B}=4\left(N_{D}-1\right)$ in the 2D case. Both the training set and boundary sets are used as the inputs of the same ANN to get their output by a feed-forward neural network.

{\romannumeral2}) The step two is to construct the loss function. To guarantee the obtained DL-FP solutions are the solutions of Eq.~\eqref{eq:3}, three conditions should be satisfied. The first is the form of Eq.~\eqref{eq:3}, the second is the normalization condition of PDF or JPDF and the third is the boundary condition. For these purposes, we define the following loss function
\begin{equation}     
	\mathcal{L}=\sum_{i=1}^{3} a_{i} \cdot E_{i}^{2}, \label{eq:9}
\end{equation}
where
\begin{equation}     
	E_{1}^{2}=a_{1} \cdot \frac{1}{N_{D}} \cdot \sum_{i=1}^{N_{D}}\left|\mathcal{N}\left(p\left(X_{i}^{D} | \theta\right)\right)\right|^{2}, \label{eq:10}
\end{equation}
\begin{equation}     
	E_{2}^{2}=a_{2} \cdot\left|\sum_{i=1}^{{ N_{D}}} \Delta t^{D i m} \cdot p\left(X_{i}^{D} | \theta\right)-1\right|^{2}, \label{eq:11}
\end{equation}
\begin{equation}     
	E_{3}^{2}=a_{3} \cdot \frac{1}{N_{B}}\left|\sum_{i=1}^{{N_{B}}} p\left(X_{i}^{B} | \theta\right)\right|^{2}. \label{eq:12}
\end{equation}
Here $a_i, i=1,2,3$ represent the penalty factors, and $Dim$ denotes dimension of the variable $X$.

We find that Eq.~\eqref{eq:10} guarantees the approximate solution to satisfy the form of Eq.~\eqref{eq:3}, Eq.~\eqref{eq:11} guarantees the normalization of PDF or JPDF, and Eq.~\eqref{eq:12} guarantees the boundary condition. These three supervisory conditions are indispensable, and the approximate solution obtained from DL-FP method will be zero if only the form of the function and the boundary condition are applied. Therefore, the normalization condition must be considered as a supervisory condition to avoid such situation that the approximate solution is zero.

{\romannumeral3}) The step three is to minimize the loss function $\mathcal{L}$. In this paper, we use the Adam method to minimize $\mathcal{L}$. We can also use the L-BFGS-B method \cite{Zhu1997Algorithm}, and a comparison between the two methods will be discussed in section 4.3. Since there are three supervisory targets in the loss function, there will be a sequence in the optimization process. In order to avoid the loss function falling into the local minimum, we design a penalty factor to affect the process of supervisory training. Through our continuous experiments, it is found that the penalty factor can be taken as one in the 1D case. In the 2D case, the normalized supervision is easy to be satisfied. On the contrary, the form of Eq.~\eqref{eq:3} is not easy to be satisfied. So we set the penalty factors as $a_1=N_D$, $a_2=0.1$, $a_3=1$. The reason why we set the rules this way is that the training set increases in the 2D case. So if the penalty factors of Eq.~\eqref{eq:10} increases, the error of each training point will be enlarged. Consequentely, the supervisory condition of the form of function will be much accounted in the minimization procedure

Algorithm \ref{alg:DL-FP} is the implemental process of the proposed DL-FP algorithm, and four different systems will be given to test the vality of our method in the next step.
\begin{algorithm}[t]
	\caption{DL-FP algorithm} 
	\hspace*{0.02in} {\bf Input:} 
	The training set $\mathcal{T}_{D}=\left\{X_{i}^{D}\right\}_{i=1}^{{N_{D}}}$ , the boundary training set $\mathcal{B}_{B}=\left\{X_{i}^{B}\right\}_{i=1}^{{N_{B}}}$, the size of ANN. \\
	\hspace*{0.02in} {\bf Output:} 
	The	DL-FP solution.
	\begin{algorithmic}[1]
		\State Initialize the weights and biases of the neural network based on the ANN size.\
		\For{iteration $j=1,2,...,\mathcal{M}$} 
		\State Construct neural network only for training set and get the output $p(X | \theta_{j})$.  \
		\State Take the first and second derivatives of the output $p_{X}(X | \theta_{j})$, $p_{X X}(X | \theta_{j})$. \
		\State Construct the loss function.
		\State Update the neural network by minimizing the loss function\
		\[
			\mathcal{L}=a_{1} \cdot \frac{1}{N_{D}} \cdot \sum_{i=1}^{N_{D}}\left|\mathcal{N}\left(p\left(x_{i}^{D} | \theta_{j} \right)\right)\right|^{2}+a_{2} \cdot\left|\sum_{i}^{N_{D}} \Delta t^{D i m} \cdot p\left(x_{i}^{D} | \theta_{j} \right)-1\right|^{2}+a_{3} \cdot \frac{1}{N_{B}} \cdot \sum_{i=1}^{N_{B}}\left|p\left(x_{i}^{B} | \theta_{j} \right)\right|^{2},
		\]
		\State Update the network:
		\[
			\theta_{j+1} \leftarrow \theta_{j} + \delta\theta_{j}
		\]
		\EndFor
		\State \Return results
	\end{algorithmic}
	\label{alg:DL-FP}%
\end{algorithm}

\section{Examples} 
The proposed framework in Section 2 provides a universal treatment of linear and nonlinear FP equations of fundamentally different nature, which are demonstrated by means of four examples including the 1D and 2D systems.

\subsection{Example 1}
Consider a nonlinear system subject to a Gaussian noise as
\begin{equation}     
	\dot{x}=\alpha x-\beta x^{3}+\sigma \Gamma,(\alpha>0, \beta>0), \label{eq:21}
\end{equation}
where $\Gamma$ is a standard Gaussian white noise. The corresponding stationary FP equation is
\begin{equation}     
	-\frac{\partial}{\partial x}\left[\left(a x-\beta x^{3}\right) p(x)\right]+\frac{\sigma^{2}}{2} \frac{\partial^{2}}{\partial x^{2}} p(x)=0. \label{eq:22}
\end{equation}
Then, its exact steady-state solution which is used to compare with the DL-FP solution is given by 
\begin{equation}     
	p_{s}(x)=C \cdot \exp \left[\frac{1}{2 \sigma^{2}}\left(2 \alpha x^{2}-\beta x^{4}\right)\right], \label{eq:23}
\end{equation}
where $C$ is the normalization constant. Using the output $p(x | \theta)$ of the deep neural network to fit $p(x)$, we can construct the loss function as follows:
\begin{equation}     
\mathcal{L}=\sum_{i=1}^{3} a_{i} \cdot E_{i}^{2}, \label{eq:27}
\end{equation}
where
\begin{equation}     
	E_{1}^{2}=a_{1} \cdot \frac{1}{N_{D}} \sum_{i=1}^{{ N_{D}}}\left|-\frac{\partial}{\partial x}\left[\left(\alpha x_{i}^{D}-\beta\left(x_{i}^{D}\right)^{2}\right) \cdot p\left(x_{i}^{D} | \theta\right)\right]+\frac{\sigma^{2}}{2} \cdot \frac{\partial^{2}}{\partial x^{2}} p\left(x_{i}^{D} | \theta\right)\right|^{2}, \label{eq:24}
\end{equation}
\begin{equation}     
	E_{2}^{2}=a_{2} \cdot \left|\sum_{i=1}^{{N_{D}}} \Delta t \cdot p\left(x_{i}^{D} | \theta\right)-1\right|^{2}, \label{eq:25}
\end{equation}
\begin{equation}     
	E_{3}^{2}=a_{3} \cdot \frac{1}{N_{B}}\left|\sum_{i=1}^{{N_{B}}} p\left(x_{i}^{B} | \theta\right)\right|^{2}. \label{eq:26}
\end{equation}

To quantify the performance and accuracy of the DL-FP algorithm, we define
\begin{equation}     
	acc_{L_{2}}=1-\frac{\|\hat{y}-y\|_{2}}{\|y\|_{2}}, \label{eq:20}
\end{equation}
where $\hat{y}$ is the approximate solution, i.e. the DL-FP solution and $y$ is the exact solution and $\|\cdot\|_{2}$ denotes the $L_2$-norm.

\begin{figure}[H]
	\centering
	\subfigure {\includegraphics[height=2in,width=2in,angle=0]{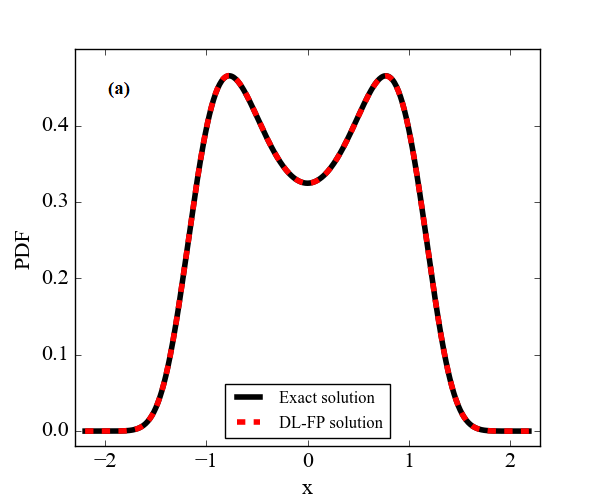}}
	\subfigure {\includegraphics[height=2in,width=2in,angle=0]{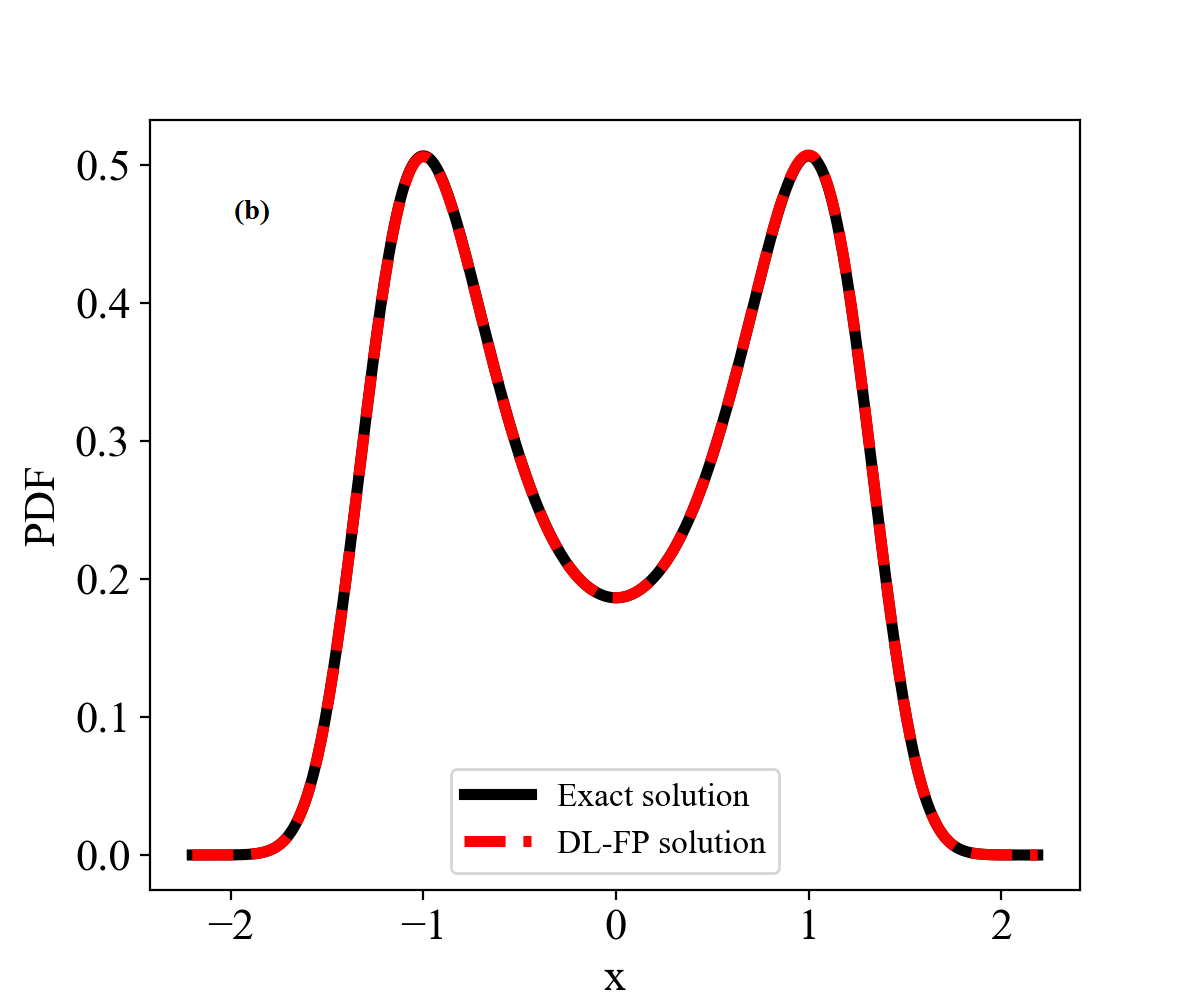}}
	\subfigure {\includegraphics[height=2in,width=2in,angle=0]{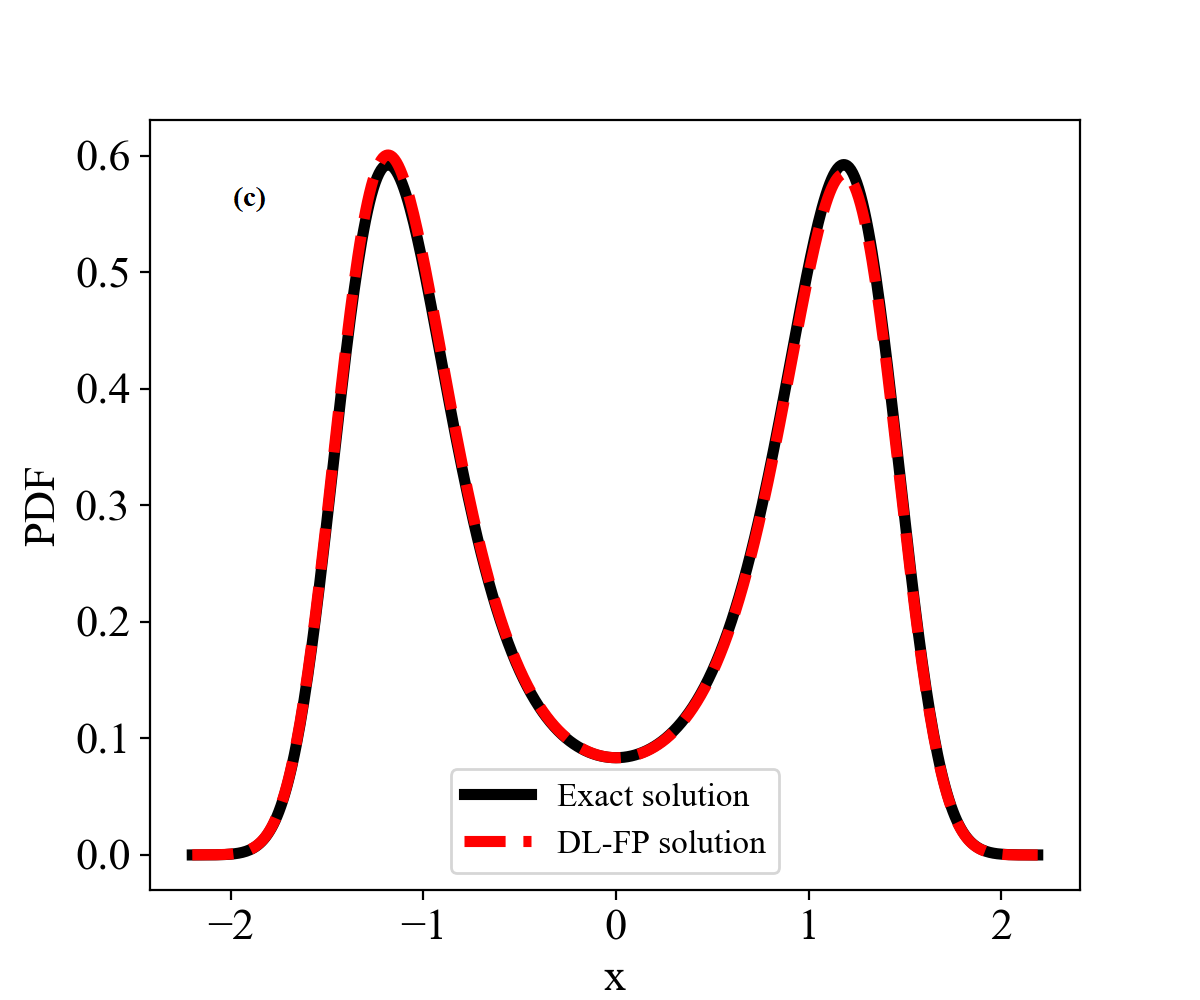}}
	\caption{The comparison results between the solution obtained from DL-FP algorithm and the exact solution for parameters $\beta=0.5$, $\sigma=0.5$. (a) $\alpha=0.3$; (b) $\alpha=0.5$; (c) $\alpha=0.7$.  The red dash-line is the DL-FP solution, and the black solid-line is the exact solution.}
	\label{fig:3}
\end{figure}

Based on the DL-FP method, we set the penalty factors to be $a_1=1$, $a_2=1$, $a_3=1$ and the training interval of the variable $x$ to be $[-2.2,2.2]$. We can get the training set and the boundary set with the step length of $\Delta t=0.01$. Setting the size of the network with four layers and 20 nodes of each hidden layer and applying the Adam optimization technique, the minimum of the loss function is about $10^{-7}$ after $3\times 10^{4}$ iterations. Fig.~\ref{fig:3} and Tab.~\ref{tab:3} show the obtained results. The red dash-line is the DL-FP solution which is almost the same as the exact solution marked by the black solid-line, and the average accuracy is 99.49$\%$. These results indicate that the DL-FP method is highly effective in solving the FP equation.

\begin{table}[H]
	\small
	\centering
	\caption{The accuracy of the DL-FP method for the system \eqref{eq:22}.}
	\begin{tabular}{llll}
		\hline
		Parameters & Loss value & $acc_{L_{2}}$ & Iterations \\
		\hline
		$\alpha=0.3$, $\beta=0.5$, $\sigma=0.5$ & $3.76 \times 10^{-7} $& 0.9995& $3\times 10^{4}$ \\		
		$\alpha=0.5$, $\beta=0.5$, $\sigma=0.5$ & $8.33 \times 10^{-7} $ & 0.9988& $3\times 10^{4}$ \\
		$\alpha=0.7$, $\beta=0.5$, $\sigma=0.5$ & $1.84 \times 10^{-7} $ & 0.9865& $3\times 10^{4}$ \\		
		\hline		
	\end{tabular}%
	\label{tab:3}%
\end{table}%

\subsection{ Example 2}
Consider the following SDE driven by an additive Gaussian noise \cite{ma2019predicting}
\begin{equation}     
	\dot{x}=f(x)+\sigma \Gamma, \label{eq:13}
\end{equation}
where $f(x)=\alpha-x+\frac{x^{8}}{1+x^{8}}$, $\Gamma$ is a standard Gaussian white noise, and $\sigma$ is the noise intensity. The corresponding stationary FP equation is
\begin{equation}     
	-\frac{\partial}{\partial x}\left[ f(x) \cdot p(x) \right]+\frac{\sigma^{2}}{2} \frac{\partial^{2}}{\partial x^{2}} p(x)=0. \label{eq:14}
\end{equation}

\begin{figure}[H]
	\centering
	\subfigure {\includegraphics[height=2in,width=2in,angle=0]{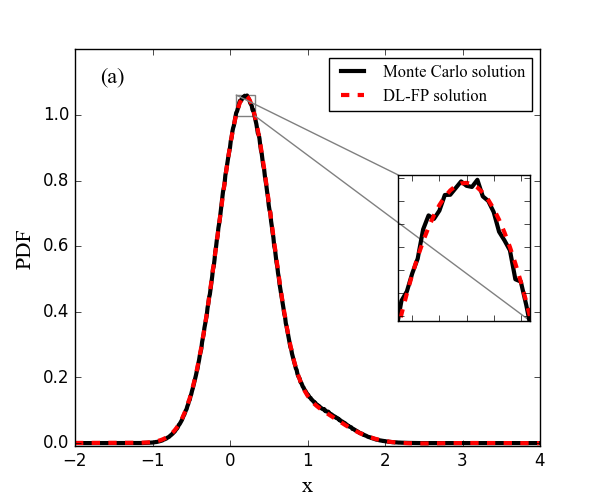}}
	\subfigure {\includegraphics[height=2in,width=2in,angle=0]{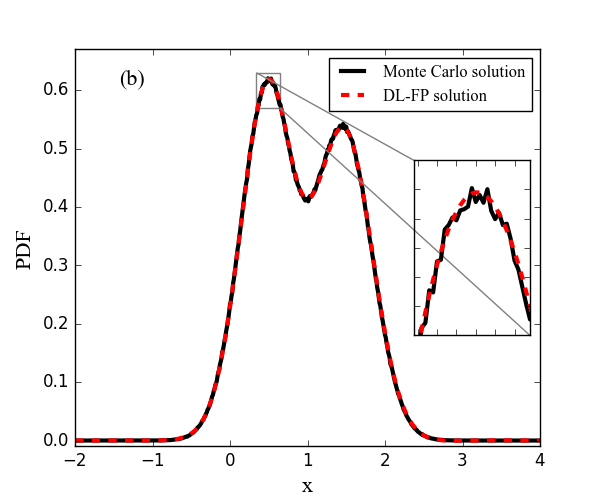}}
	\subfigure  {\includegraphics[height=2in,width=2in,angle=0]{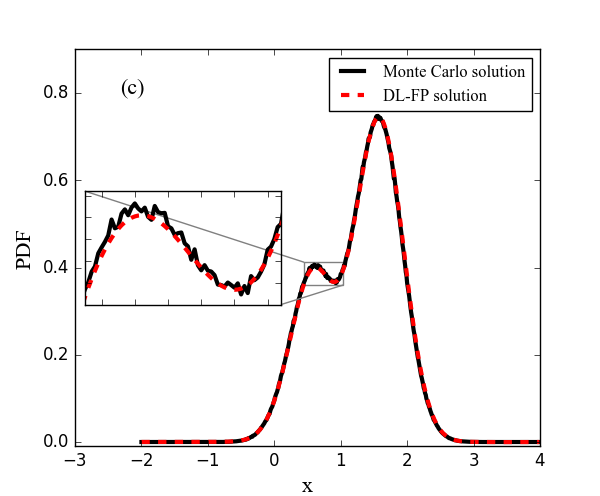}}
	\caption{The comparison results between the solutions obtained from the DL-FP algorithm and the Monte Carlo method with fixed noise intensity $\sigma=0.5$ and different parameter for the system \eqref{eq:14}. (a) $\alpha=0.2$; (b) $\alpha=0.5$; (c) $\alpha=0.6$. The red  dash-line denotes the DL-FP solution, and the black solid-line for the Monte Carlo solution.}
	\label{fig:2}
\end{figure}

In order to check the validity of the DL-FP method, the Monte Carlo solutions are also given which are shown in Fig.~\ref{fig:2}.

In the calculation, the training interval of the variable $x$ is selected to be $[-2,4]$, then $p( -2)=0$ and $p(4)=0$. Setting the step length to be $\Delta t=0.01$ to get the training set and the boundary set is $\left\{-2,4\right\}$. With the DL-FP algorithm outlined above, the loss function is then minimized with deep nerual network which owns four layers and having 20 hidden nodes in every hidden layer. The activation function is selected as the tanh function, using the Adam method to minimize the loss function and the minimum is about $10^{-6}$ after $5 \times 10^{4}$ iterations. Monte Carlo simulation is applied to get the PDF of Eq.~\eqref{eq:13}. We adopt the second-order stochastic Runge-Kutta algorithm \cite{Honeycutt1992Stochastic} to generate $10^{8}$ data, then carry on the statistics to get the results which are shown in the black solid-line in Fig.~\ref{fig:2}. Compared with the DL-FP solution, we find that the Monte Carlo solution is consistent with the DL-FP one. However, after zooming some loacal details, in contrast with the roughness of the Monte Carlo solution, the DL-FP result is rather smooth.

\subsection{Example 3}
The Duffing oscillator can be described as
\begin{equation}     
\ddot{x}+\eta x+\alpha x+\beta x^{3}=\sigma \Gamma, \label{eq:28}
\end{equation}
where $\eta$ is the damping coefficient, $\alpha$ and $\beta$ denote the linear and nonlinear stiffness respectively, $\Gamma$ represents a standard Gaussian white noise with $\sigma$ the noise intensity. Let
\begin{equation}     
\left\{\begin{array}{c}{\dot{x}=y,} \\ {\dot{y}=-\eta y+\alpha x+\beta x^{3}+\sigma \Gamma.}\end{array}\right. \label{eq:29}
\end{equation}
Then the corresponding stationary FP equation is
\begin{equation}     
	-\frac{\partial}{\partial x}[y p(x, y)]-\frac{\partial}{\partial y}\left[\left(-\eta y-\alpha x-\beta x^{3}\right) p(x, y)\right]+\frac{\sigma^{2}}{2} \frac{\partial^{2}}{\partial y^{2}} p(x, y)=0. \label{eq:30}
\end{equation}
and the exact steady-state solution is
\begin{equation}     
	p_{s}(x, y)=C \cdot \exp \left[-\frac{\eta}{\sigma^{2}}\left(y^{2}+\alpha x^{2}+\frac{\beta}{2} x^{4}\right)\right], \label{eq:31}
\end{equation}
where $C$ is the normalization constant.

\begin{figure}[H]
	\centering
	\subfigure {\includegraphics[height=2in,width=2.1in,angle=0]{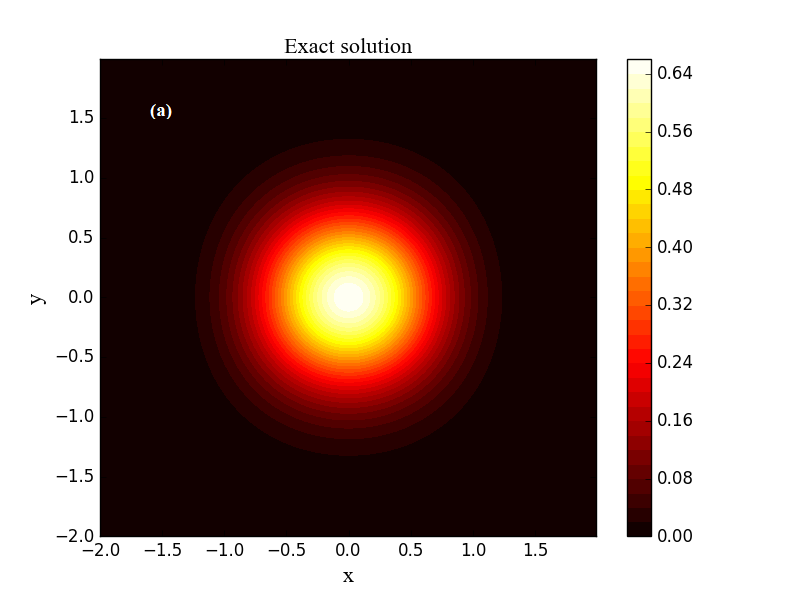}}
	\subfigure  {\includegraphics[height=2in,width=2.1in,angle=0]{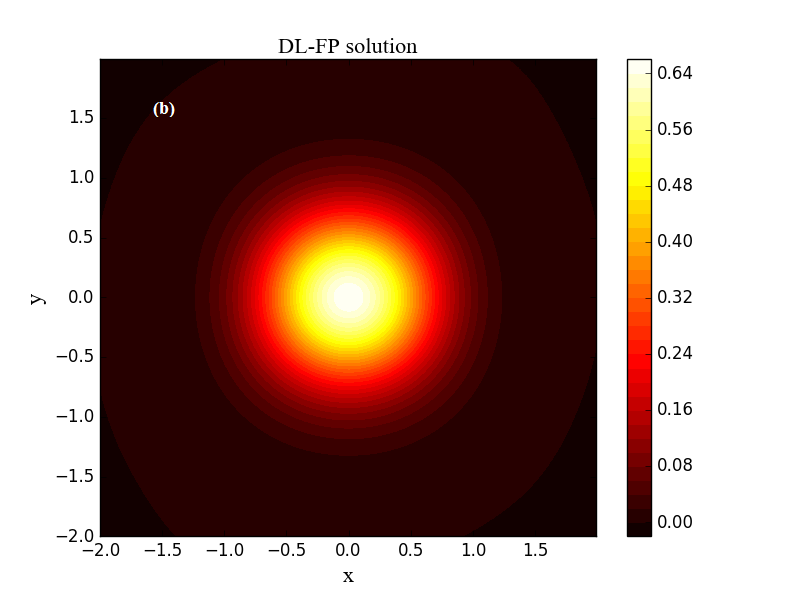}}
	\subfigure  {\includegraphics[height=2in,width=2.1in,angle=0]{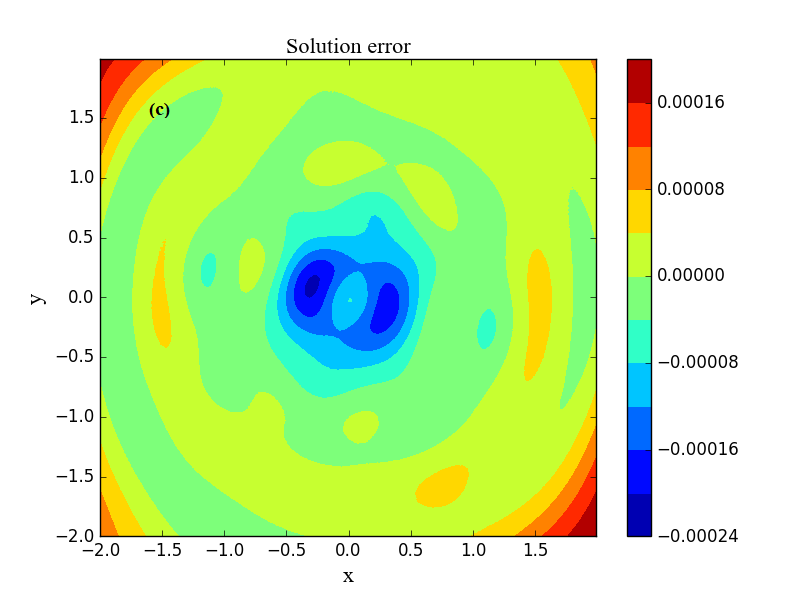}}
	
	\subfigure {\includegraphics[height=2in,width=2in,angle=0]{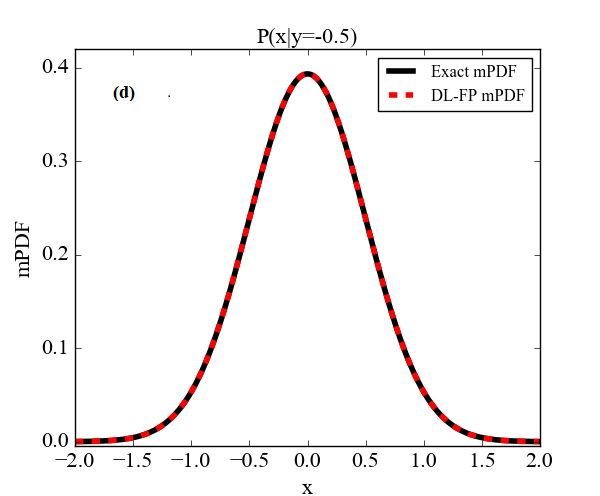}}
	\subfigure  {\includegraphics[height=2in,width=2in,angle=0]{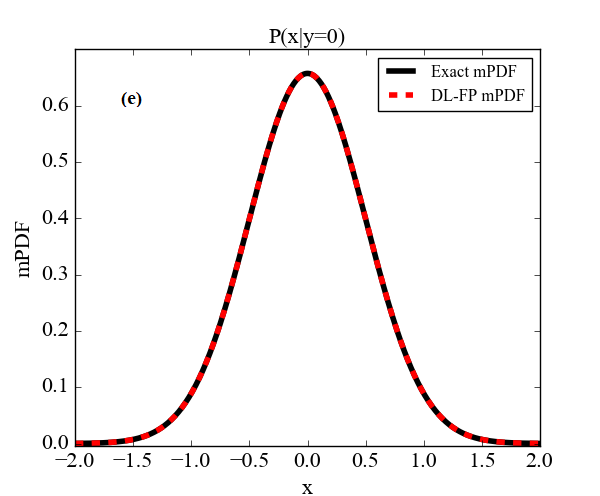}}
	\subfigure  {\includegraphics[height=2in,width=2in,angle=0]{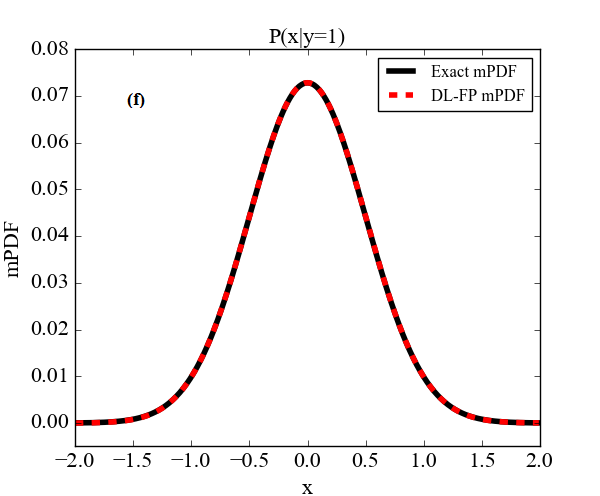}}
	\caption{The comparison results for the the Duffing system \eqref{eq:28} with: $\alpha=1.0$, $\beta=0.2$, $\eta=0.2$, $\sigma^2=0.1$. (a) The contour map of the exact JPDF; (b) The contour map of the DL-FP JPDF; (c) The error between the DL-FP solution and the exact solution; (d) The contrast between the exact marginal probability density (mPDF) and the DL-FP mPDF when $y=-0.5$; (e) The contrast between the exact mPDF and the DL-FP mPDF when $y=0$; (f) The contrast between the exact mPDF and the DL-FP mPDF when $y=1$.}
	\label{fig:4}
\end{figure}

Minimizing the loss function using the Adam optimization algorithm with $5 \times 10^{4}$ iterations, we get the results as is shown in Fig.~\ref{fig:4}. Here, we set in \eqref{eq:30} to be $\alpha=1.0$, $\beta=0.2$, $\eta=0.2$, $\sigma^2=0.1$. The inputs of the ANN is the discretization using the step of 0.01 with the range of $[-2,2] \times[-2,2]$. The training set owns $1.6 \times 10^{5}$ data and the boundary set owns $1596$ data. The boundary conditions are assumed to be zero for all the states.

Using the DL-FP method to solve the Eq.~\eqref{eq:30}, we set the penalty factors as $a_1=400$, $a_2=0.1$, $a_3=1$ based on the rules of the DL-FP method in 2D case, and set the network with four hidden layers with 20 nodes in each hidden layer.  The DL-FP solution is almost the same as the exact solution where the maximum absolute error in the JPDF is  at the order $10^{-4}$. The accuracy of the DL-FP solution can be seen more intuitively through the marginal probability density function (mPDF) as indicated in Fig.~\ref{fig:4} (d,e,f). This example shows the vality of this algorithm in dealing with a single-degree-of-freedom Duffing system.

\subsection{Example 4}
Consider a Van der Pol system as
\begin{equation}     
\ddot{x}+\gamma\left(-1+x^{2}+\dot{x}^{2}\right) \dot{x}+x=\sigma \Gamma, \label{eq:32}
\end{equation}
where $\gamma$ is a constant, $\Gamma$ is a standard Gaussian white noise, and $\sigma$ denotes the noise intensity. We let
\begin{equation}     
\left\{\begin{array}{c}{\dot{x}=y,} \\ {\dot{y}=-\gamma\left(-1+x^{2}+y^{2}\right) y-x+\sigma \Gamma(t).}\end{array}\right. \label{eq:33}
\end{equation}
Then one can get the reduced FP equation
\begin{equation}     
-\frac{\partial}{\partial x}[y p(x, y)]-\frac{\partial}{\partial y}\left\{\left[-\gamma\left(-1+x^{2}+y^{2}\right) y+x\right] p(x, y)\right\}+\frac{\sigma^{2}}{2} \frac{\partial^{2}}{\partial y^{2}} p(x, y)=0, \label{eq:34}
\end{equation}
 and the corresponding exact steady-state solution
\begin{equation}     
	p_{s}(x, y)=C \cdot \exp \left\{\frac{\gamma}{\sigma^{2}}\left[\left(x^{2}+y^{2}\right)-\frac{1}{2}\left(x^{2}+y^{2}\right)^{2}\right]\right\}, \label{eq:35}
\end{equation}
where $C$ is the normalization constant.

\begin{figure}[H]
	\centering
	\subfigure {\includegraphics[height=2in,width=2.1in,angle=0]{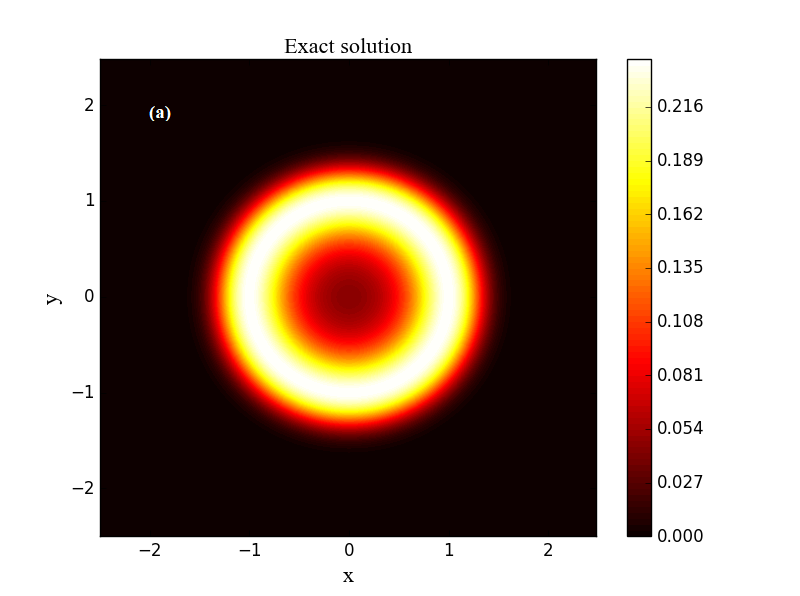}}
	\subfigure  {\includegraphics[height=2in,width=2.1in,angle=0]{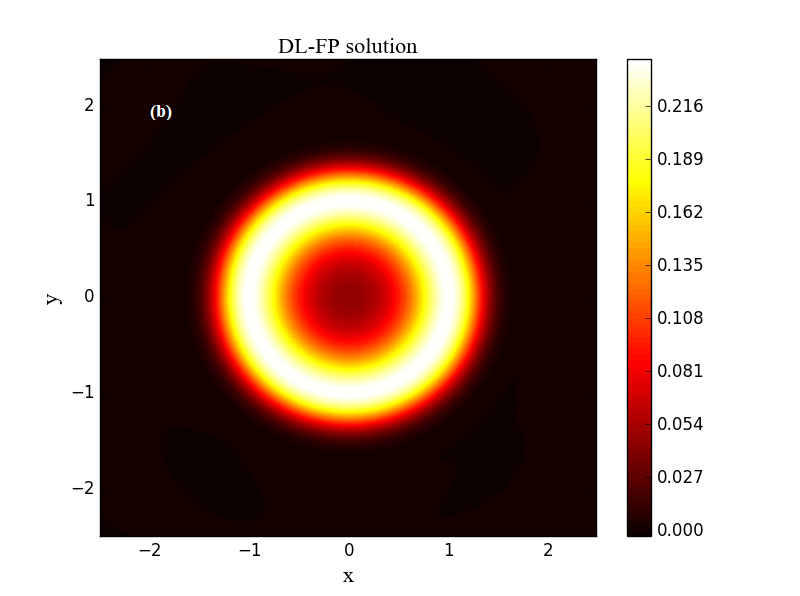}}
	\subfigure  {\includegraphics[height=2in,width=2.1in,angle=0]{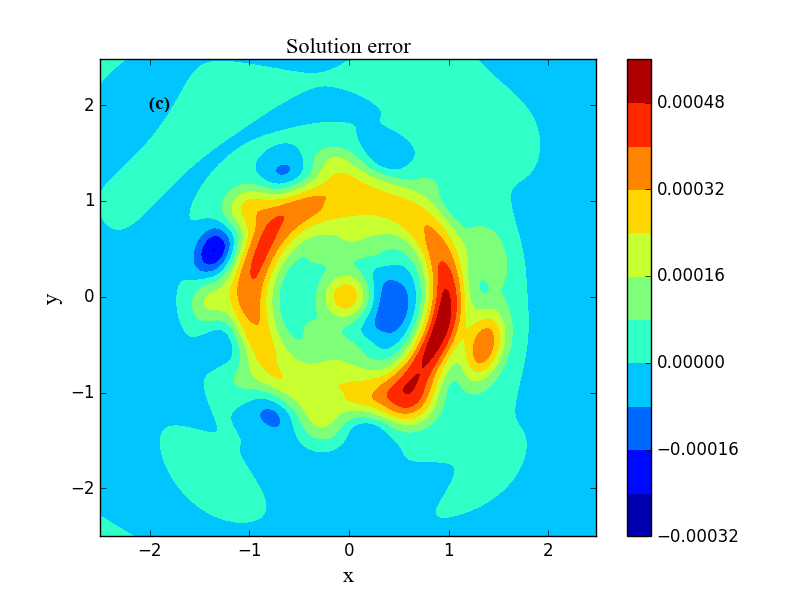}}
	
	\subfigure {\includegraphics[height=2in,width=2in,angle=0]{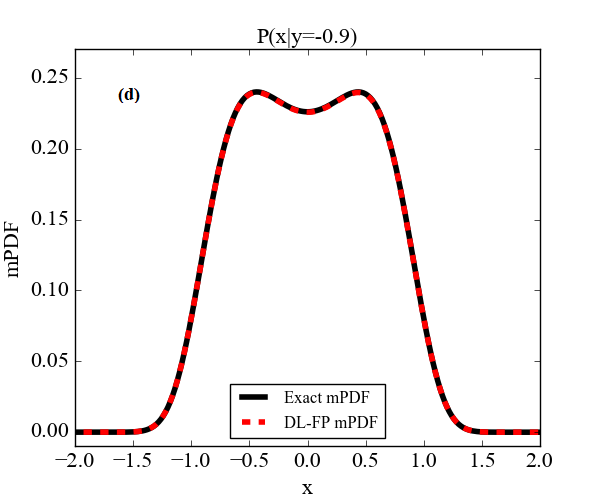}}
	\subfigure  {\includegraphics[height=2in,width=2in,angle=0]{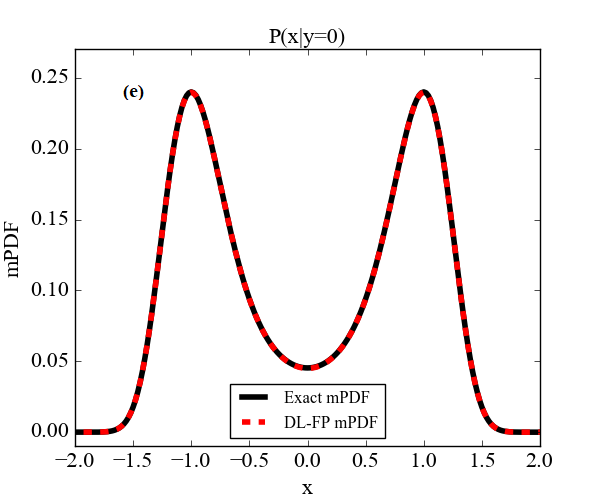}}
	\subfigure  {\includegraphics[height=2in,width=2in,angle=0]{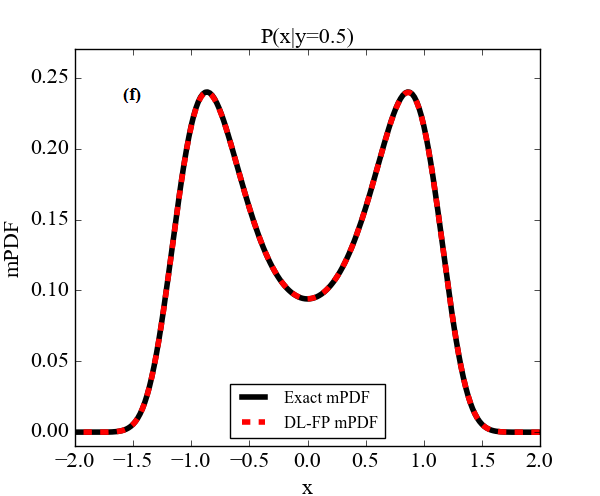}}
	\caption{The comparison results for the Van-der-Pol system \eqref{eq:32} with: $\alpha=1.0$, $\beta=0.2$, $\eta=0.2$, $\sigma^2=0.1$. (a) The contour map of the exact JPDF; (b) The contour map of the DL-FP JPDF; (c) The error between the DL-FP solution and the exact solution; (d) The contrast between the exact mPDF and the DL-FP mPDF when $y=-0.9$; (e) The contrast between the exact mPDF and the DL-FP mPDF when $y=0$; (f) The contrast between the exact mPDF and the DL-FP mPDF when $y=0.5$.}
	\label{fig:5}
\end{figure}

Based on the DL-FP method, we solve Eq.~\eqref{eq:34} with $\gamma=1.0$, $\sigma^2=0.3$ . Step 1, selects the training set with the range of $[-2,2] \times[-2,2]$, then discretize the interval via the step of 0.01 to get $1.6 \times 10^{5}$ data in the training set and also get 1596 data in the boundary set. Step 2, sets the network with four hidden layers with 20 nodes in each hidden layer, and the penalty factor as $a_1=400$, $a_2=0.1$, $a_3=1$ based on the rules of the 2D situation. Initializing the weight and bias of the neural network based on the ANN size and minimizing the loss function by the Adam optimization algorithm with $5 \times 10^{4}$ iterations, we present the results in Fig.~\ref{fig:5}. 

We uncover that the DL-FP solutions agree well with the exact solutions as shown in Fig.~\ref{fig:5} (a,b). the solution error is shown in Fig.~\ref{fig:5}(c), the maximum absolute error in JPDF is at the order $10^{-4}$, and it is satisfactory that the error in the tail is trained well. We also observe the accuracy of the DL-FP method through the mPDF in Fig.~\ref{fig:5}(d,e,f).

\section{Discussion}
The DL-FP algorithm is proposed in this paper, and can be observed the high accuracy of the DL-FP algorithm in Section 3 due to the particularity of the FP equation and the setting up of the network. In this section, we will discuss three problems which are concerned in the proposed DL-FP algorithm: (1) The effect of the number of hidden layers on the calculated performances; (2) The importance of the penalty factors in the algorithm; (3) How to select the optimization algorithm to train the FP equation.

\subsection{The number of hidden layers}
The influence of the network size on the accuracy is discussed firstly. For the selection of the optimization algorithm, then the Adam optimization algorithm is applied to illustrate the effect of network size on the accuracy. We will discuss the 1D and 2D systems respectively.

\begin{figure}[H]
	\centering
	\subfigure {\includegraphics[height=3in,width=3in,angle=0]{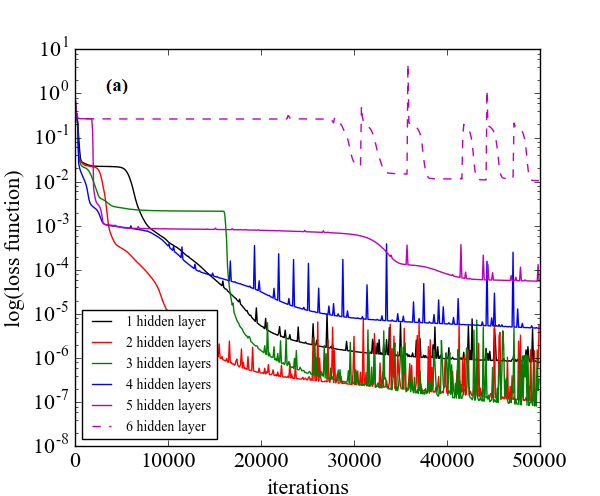}}
	\subfigure {\includegraphics[height=3in,width=3in,angle=0]{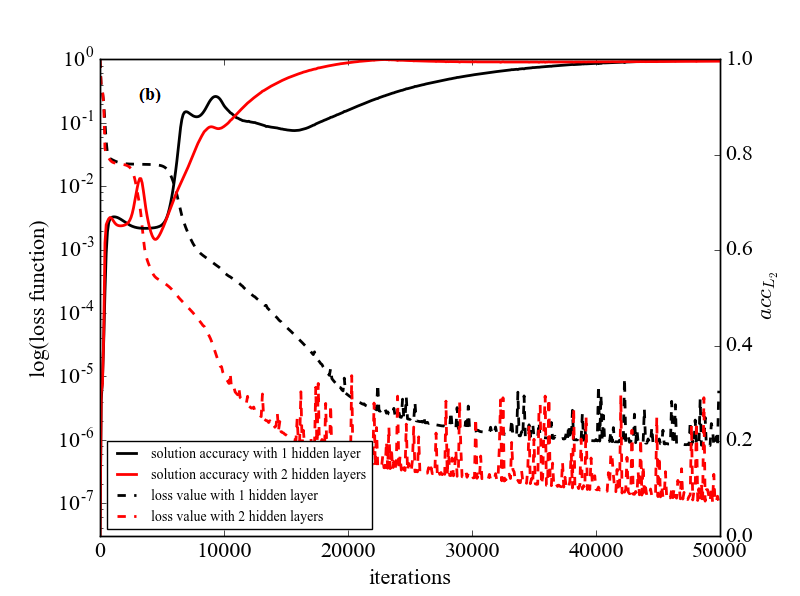}}
	\caption{(a) The comparison of loss functions for increasing network depth in terms of the system \eqref{eq:21}; (b) The accuracy varies with the loss function in double-well system (The solid line represents the value of the loss function which corresponds to the left axis, and the dotted line represents the accuracy which corresponds to the right axis). }
	\label{fig:6}
\end{figure}

\begin{figure}[H]
	\centering
	\subfigure {\includegraphics[height=3in,width=3in,angle=0]{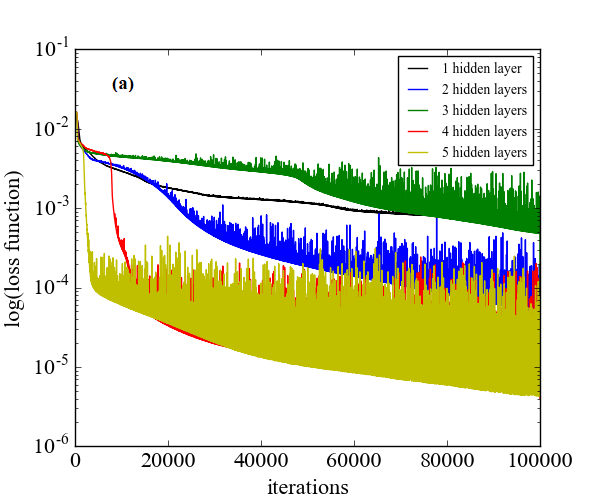}}
	\subfigure  {\includegraphics[height=3in,width=3in,angle=0]{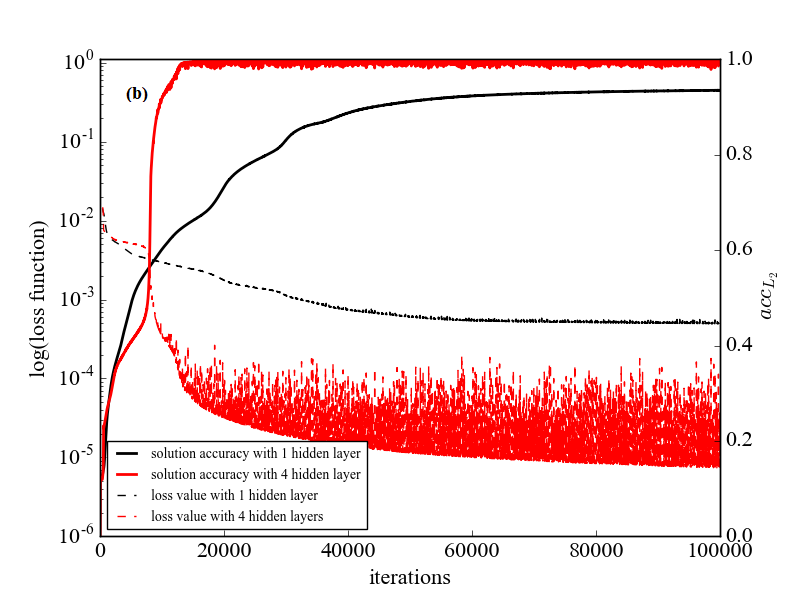}}
	\caption{ (a) The comparison of loss function for increasing network depth for the Duffing system \eqref{eq:23}; (b) The accuracy varies with the loss function in \eqref{eq:23} (The solid line represents the value of the loss function which corresponds to the left axis, and the dotted line represents the accuracy which corresponds to the right axis). }
	\label{fig:7}
\end{figure}

For the 1D system, we consider the system \eqref{eq:21} with fixed $\alpha=0.3$, $\beta=0.5$, $\sigma=0.5$ as an example, where the network is assumed to have 1, 2, 3, 4, 5 and 6 hidden layers with 20 nodes in each hidden layer. We clearly observe in Fig.~\ref{fig:6} that the loss function will reach the order of $10^{-7}$ when the hidden layer is 2 and 3 which is better than a single hidden layer in Fig.~\ref{fig:7} (a). However, the trend of the loss function gets worse when the number of the hidden layers exceeds 4. Especially, when the number of the hidden layer is 6, it's difficult to train the network as $\mathcal{L}$ falling into the local optimum. So the best number of hidden layer is 2 or 3 with 20 nodes in each layer. From Fig.~\ref{fig:6} (b) we can find that 2 hidden layers do not significantly improve the final accuracy of the calculation after a large number of iterations. Both 1 and 2 hidden layers are able to keep value of $acc_{L_{2}}$ at around 0.996. However, more hidden layers do increase the speed to reach a high accuracy. A single hidden layer can be used to obtain the higher accuracy under the Adam optimization algorithm since that a deeper hidden layer is unnecessary.

For the 2D system, we take the Duffing system with $\alpha=1.0$, $\beta=0.2$, $\eta=0.2$, $\sigma^2=0.1$ as an example and use the networks with 1, 2, 3, 4 and 5 hidden layers with 20 nodes in each hidden layer. The results are given in Fig.~\ref{fig:7}. Different from the 1D case, the influence of more hidden layers on the accuracy of calculation and the optimization are more advantageous than a single hidden layer. In Fig.~\ref{fig:7} (a), with the increase of the number of hidden layers, the loss function decreases more rapidly. When there are 5 hidden layers, it will quickly reach the order of $10^{-5}$, while the single hidden layer can only reach the order of $10^{-3}$. The multi-hidden layers work better for the improvement of accuracy in Fig.~\ref{fig:7} (b). The accuracy of a single hidden layer is maintained at about 0.94, while the 4 hidden layers can reach about 0.99. So, in 2D computation, it is very helpful to use more hidden layers to improve the speed of training and the accuracy.

\subsection{The penalty factor}
This section is to discuss the effect of a penalty factor in the 2D case and take the  Duffing system eq.~\eqref{eq:23} as an example. We select the interval to be $[-2,2] \times[-2,2]$ and $\Delta t=0.05$,  $\alpha=1.0$, $\beta=0.2$, $\eta=0.2$, $\sigma^2=0.1$. We use an ANN with 5 hidden layers where the nodes in each hidden layer are 20.  

\begin{figure}[H]
	\centering
	\subfigure {\includegraphics[height=3in,width=3.2in,angle=0]{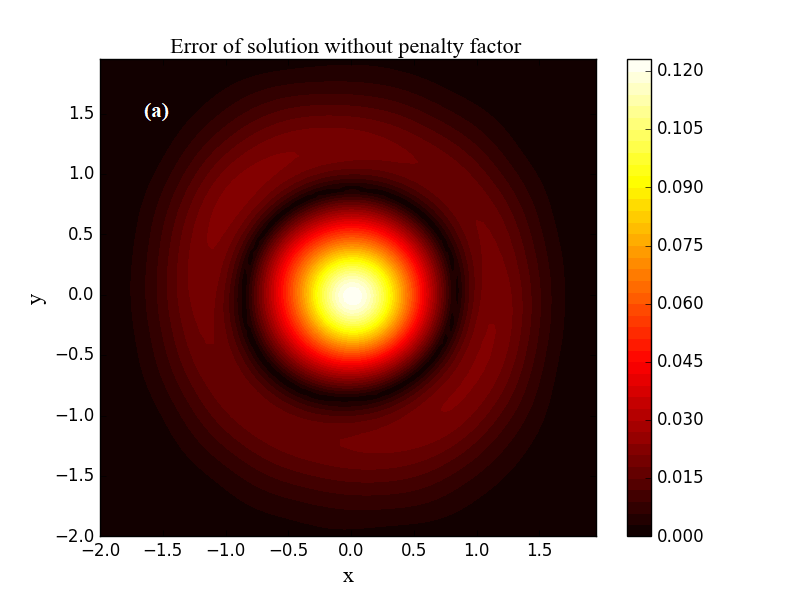}}
	\subfigure  {\includegraphics[height=3in,width=3.2in,angle=0]{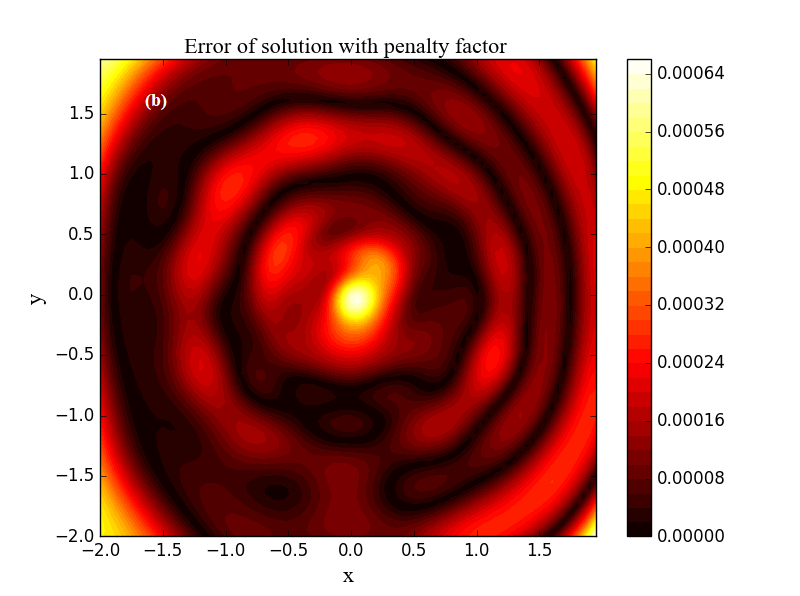}}
	\caption{Influences of the penalty factors on the absolute error of solution for the Duffing system \eqref{eq:23}. (a) without penalty factors; (b) with penalty factors.}
	\label{fig:8}
\end{figure}

We set the penalty factor to be $a_1=1$, $a_2=1$, $a_3=1$ which is equal to the traditional mean square loss function. After training $1 \times 10^{5}$ times, the result of the loss function is about $1.52 \times10^{-7}$, and the absolute error between the DL-FP solution and the exact solution is shown in Fig.~\ref{fig:8} (a). We observe that the solution error is very small at the boundary, but it is large in the middle part. Because the trial solution is easy to satisfy the boundary condition and the supervision training on the form of the equation is not strict enough, which lead to the training failure. So, we increase the punishment of Eq.~\eqref{eq:10}. According to the rules in section 2, we set the penalty factor to be $a_1=80$, $a_2=0.1$, $a_3=1$. After training $1 \times 10^{5}$ times, the result of loss function is $1.52 \times10^{-6}$ and the absolute error between the DL-FP and the exact solutions is shown in Fig.~\ref{fig:8} (b). Hence, the solution error is about the order of $10^{-4}$. When the weight of Eq.~\eqref{eq:10} increases, the error caused by each data point in the training will be enlarged. Besides, the boundary condition is easily satisfied, so that the training is successful. Therefore, it is very necessary to add the penalty factor in the loss function.

\subsection{The optimization algorithm}
Selecting the appropriate optimization algorithm is very important during training the network. The Adam optimization algorithm is adopted in the discussion of section 4.1. It was found that a relatively good training can only be achieved when the iteration reaches $5 \times 10^{4}$, and the training time of this method will reach one hour in the 2D case which is too long to be accepted for a fast operation. Hence, we try to improve the optimization algorithm. It is easy to find that the gradient disappears if the stochastic gradient descent method and momentum method are adopted. We find that the L-BFGS-B algorithm \cite{Zhu1997Algorithm} of the quasi-Newton method is very effective. We compare the calculation of the Adam and the L-BFGS-B methods in the 1D and 2D cases and the results are shown in Tab.~\ref{tab:4} and Tab.~\ref{tab:5}. All the tests are carried out in a GPU mode on a computer with a video card of Quadro k620.  

Firstly, we take the system \eqref{eq:21} as an example. Set the interval to be $[-2.3,2.3]$ and the parameters to be $\Delta t=0.01$, $\alpha=0.3$, $\beta=0.5$, $\sigma=0.5$. Both methods set the number of iteration to be $5 \times 10^{4}$ and the number of nodes in each layer of the network to be 20. The results of the comparison of the calculation with different numbers of hidden layers are shown in Tab.~\ref{tab:4}. We find that the L-BFGS-B method has a significant kicking effect on the calculation time which can ensure that the calculation can be completed in about 30s, and the effect of multi-hidden layers is better than that of the single hidden layer. Additionally, the accuracy of the training is better when the value of $\mathcal{L}$ reaches the order of $10^{-7}$ both in the Adam and the L-BFGS-B. The Adam algorithm will overfit with the increase of the number of hidden layers, but the L-BFGS-B algorithm will not.
\begin{table}[H]
	\small
	\centering
	\caption{The comparison of Adam and L-BFGS-G algorithms in the 1D case}
	\begin{tabular}{llllll}
		\hline
		Number of hidden layers & Method & Loss value & $acc_{L_{2}}$ & Maximum absolute error & Time cost \\
		\hline
		1&Adam & $2.96 \times10^{-6}$ & 0.9988 & $6.7 \times10^{-4}$& 47.2s \\
		&L-BFGS-B & $4.38 \times10^{-6}$& 0.961& $2.12 \times10^{-2}$& 33.7s \\
		\hline
		2&Adam & $1.03 \times10^{-7}$& 0.9962& $1.09 \times10^{-3}$& 78.4s \\
		&L-BFGS-B & $2.56 \times10^{-7}$& 0.9989& $5.6 \times10^{-4}$& 26.8s \\
		\hline
		3&Adam & $7.93 \times10^{-8}$& 0.9979& $1.09 \times10^{-3}$& 110.6s \\
		&L-BFGS-B & $9.03 \times10^{-8}$& 0.9971& $1.54 \times10^{-3}$& 32.9s \\
		\hline
		4&Adam & $4.61 \times10^{-6}$& 0.9368& $3.4 \times10^{-2}$& 134.6s \\
		&L-BFGS-B & $5.63 \times10^{-7}$& 0.9976& $1.28 \times10^{-3}$& 25.7s \\
		\hline
		
	\end{tabular}%
	\label{tab:4}%
\end{table}%

In the 2D case, we take the system \eqref{eq:23} as a study example. Set the interval to be $[-2,2] \times[-2,2]$ and the parameters to be $\Delta t=0.05$, $\alpha=1.0$, $\beta=0.2$, $\eta=0.2$, $\sigma^2=0.1$. The network is set up in the same way as the 1D case. It can be clearly seen in Tab.~\ref{tab:5} that the L-BFGS-B algorithm has great advantages concering computing time. Moreover, when the value of the loss function reaches the order of $10^{-6}$, the training accuracy is very high. Whether using the Adam or the L-BFGS-B method, the accuracy will increase with the number of layers. Therefore, the L-BFGS-B technique can achieve the same effect as the Adam method, but saves a lot of time.
\begin{table}[H]
	\small
	\centering
	\caption{The comparison of Adam and L-BFGS-B algorithms in the 2D case}
	\begin{tabular}{llllll}
		\hline
		Number of hidden layers & Method & Loss value & $acc_{L_{2}}$ & Maximum absolute error & Time cost \\
		\hline
		1&Adam & $6.60 \times10^{-4}$& 0.9056& $7.69 \times10^{-2}$& 1249.9s \\
		&L-BFGS-B & $2.49 \times10^{-3}$& 0.7134& $2.36 \times10^{-1}$& 149.7s \\
		\hline
		2&Adam & $4.89 \times10^{-5}$& 0.9907& $7.21 \times10^{-3}$& 2343.5s \\
		&L-BFGS-B & $4.98 \times10^{-6}$& 0.9980& $4.46 \times10^{-3}$& 240.4s \\
		\hline
		3&Adam & $1.27 \times10^{-5}$& 0.9977& $3.36 \times10^{-3}$& 3054.4s \\
		&L-BFGS-B & $9.92 \times10^{-6}$& 0.9974& $5.88 \times10^{-3}$& 183.1s \\
		\hline
		4&Adam & $4.00 \times10^{-6}$& 0.9993& $1.7 \times10^{-2}$& 3985.3s \\
		&L-BFGS-B & $5.90 \times10^{-6}$& 0.9978& $5.26 \times10^{-3}$& 264.4s \\
		\hline
		5&Adam & $4.27 \times10^{-6}$& 0.9993& $6.45 \times10^{-4}$& 4793.6s \\
		&L-BFGS-B & $1.27 \times10^{-6}$& 0.9985& $3.42 \times10^{-3}$& 353.0s \\
		\hline
		
	\end{tabular}%
	\label{tab:5}%
\end{table}%

\section{Conclusions}
This paper proposed a numerical method named DL-FP to solve the FP equation using deep learning. Four examples of 1D and 2D systems have illustrated that the proposed DL-FP has a high accuracy. Unlike traditional numerical algorithms, DL-FP can solve the continuous solution of the FP equation.
Inspired by the traditional solution to differential equations by ANNs, in this paper, the normalization condition is introduced as a supervision condition, so as to avoid the situation that the training solution of FP equation is zero.

Due to the introduction of normalized monitoring conditions, a penalty factor is given for the loss function (using the traditional mean square error format) effectively trained to minimize the calculation. We introduce a penalty factor to solve this problem effectively. In particular, this paper gives rules to set the penalty factors in 1D and 2D cases. In Section 4.2, the Duffing system \eqref{eq:23} is adopted as an example. However, the setting rules of such penalty factors still need to be studied in higher dimensions.

Besides, we discuss the setting of the number of hidden layers and the optimization algorithm in ANN. In the setting of the network, we suggest to choose deeper hidden layers to the higher accuracy. In terms of the number of iterations needed for training, we suggest that it is more reasonable to determine the training number according to the precision of the loss function. In the 1D case, the loss function reaches the order of $10^{-7}$, and in the 2D case, it reaches the order of $10^{-6}$. In the selection of algorithms for optimizing loss functions, we suggest the Adam and the L-BFGS-B algorithms. In the 1D case, there is little difference between them, but in the 2D case, the L-BFGS-B algorithm strongly shortens the training time.

\section*{Acknowledgements}
This work was partly supported by the National Natural Science Foundation of China (Grant No.11572247 \& 11772255), the Fundamental Research Funds for the Central Universities,
Shaanxi Project for Distinguished Young Scholars, and the Research Funds for Interdisciplinary subject, NWPU.

\section*{References}
\bibliography{mybibfiles}

\end{document}